\documentclass[preprint,12pt]{elsarticle}

\journal{Icarus}

\usepackage{natbib}
\usepackage{aas_macros,amsmath}
\usepackage{url}
\usepackage{array}

\usepackage{color,soul}
\newcommand{\revised}[1]{{#1}}
\newcommand{\rerevised}[1]{{#1}}
\newcommand{\rererevised}[1]{{#1}}
\usepackage{textcomp}
\biboptions{authoryear}

\begin{document}
\begin{frontmatter}

\title{Delivery of organics to Mars through asteroid and comet impacts}

\author[sronaddress,kapteynaddress]{Kateryna~Frantseva\corref{mycorrespondingauthor}}
\cortext[mycorrespondingauthor]{Corresponding author}
\ead{k.frantseva@sron.nl}

\author[kapteynaddress,sronaddress]{Michael~Mueller}

\author[utrechtaddress]{Inge~Loes~ten~Kate}

\author[sronaddress,kapteynaddress]{Floris~F.S.~van~der~Tak}

\author[caaddress1,caaddress2]{Sarah~Greenstreet}

\address[sronaddress]{SRON Netherlands Institute for Space Research, Landleven 12, 9747 AD Groningen, The Netherlands}
\address[kapteynaddress]{Kapteyn Astronomical Institute, University of Groningen, Landleven 12, 9747 AD Groningen, The Netherlands}
\address[utrechtaddress]{Department of Earth Sciences, Utrecht University, Budapestlaan 4, 3584 CD Utrecht, The Netherlands}
\address[caaddress1]{Las Cumbres Observatory, 6740 Cortona Dr., Suite 102, Goleta, CA, USA}
\address[caaddress2]{University of California at Santa Barbara, Santa Barbara, CA, USA}

\begin{abstract}

Given rapid photodissociation and photodegradation, the recently discovered organics in the Martian subsurface and atmosphere were probably delivered in geologically recent times. Possible parent bodies are C-type asteroids, comets, and interplanetary dust particles (IDPs). 

%%% MM 2018/02/19: shortened IDP part, that's not the focus of our paper
\rerevised{The dust infall rate was estimated, using different methods, to be between $0.71$ and $2.96 \times 10^6$ kg/yr }\citep{Nesvorny2011,Borin2017,Crismani2017};
%\rerevised{The total dust infall rate on Mars based on observations of IDP influx on Earth, MAVEN observations of the products of atmospheric ablation on Mars, and dynamical models is estimated to be between $0.71$ and $2.96 \times 10^6$ kg/yr} \citep{Nesvorny2011,Borin2017,Crismani2017}
\rerevised{assuming a carbon content of 10\% }\citep{Flynn1996}\rerevised{, this implies an IDP carbon flux of $0.07 - 0.3 \times 10^6$ kg/yr.}
%\revised{The IDP infall rate on Mars was estimated most recently by} \citet{Borin2017} \revised{ based on a \rerevised{dynamical} model, and by} \citet{Crismani2017} \revised{based on MAVEN observations of the products of atmospheric ablation.  Assuming a carbon content of 10\% by mass} \citep{Flynn1996}\revised{, their results imply a carbon infall rate due to IDPs of $0.3\times 10^6$~kg/yr and 0.07--$0.1 \times 10^6$ kg/yr, respectively.} 
We calculate for the first time the carbon flux from impacts of asteroids and comets.

\revised{To this end, we perform} \rerevised{dynamical simulations} of impact rates on Mars.
We use the N-body integrator {\tt RMVS/Swifter} to propagate the Sun and the eight planets from their current positions. We separately add comets and asteroids to the simulations as massless test particles, based on their current orbital elements, \revised{yielding Mars impact rates of}
\revised{$4.34\times10^{-3}$ comets/Myr and 3.3 asteroids/Myr}.

We estimate the delivered amount of carbon using published carbon content values. In asteroids, only C types contain appreciable amounts of carbon. Given the absence of direct taxonomic information on the Mars impactors, we base ourselves on the measured distribution of taxonomic types in combination with dynamic models of the origin of Mars-crossing asteroids.

We estimate the global carbon flux on Mars from cometary impacts to be  \revised{$\sim 0.013 \times 10^{6}$~kg/yr within an order of magnitude, while asteroids deliver $\sim 0.05 \times 10^6$~kg/yr.} These values correspond to \revised{$\sim 4-19 \%$ and $\sim 17-71 \%$}, respectively, of the IDP-borne carbon flux estimated by\rerevised{ }\citeauthor{Nesvorny2011}\rerevised{, }\citeauthor{Borin2017} and \citeauthor{Crismani2017}.
\revised{Unlike} the spatially homogeneous IDP infall, impact ejecta are distributed locally, concentrated around the impact site.
We find organics from asteroids and comets to dominate over IDP-borne organics at distances up to \revised{150~km} from the crater center. Our results may be important for the interpretation of \emph{in situ} detections of organics on Mars.
\end{abstract}


\end{frontmatter}

%\linenumbers

\section{Introduction}\label{introduction}

Organic molecules, made primarily of carbon, form the building blocks of life. It has therefore been a primary goal of Mars research to establish whether or not organics are present on Mars, or whether they were in the past when Mars was more hospitable to life. \revised{On the surface of Mars, organics would be expected to photodissociate within hours} \citep{tenKate2005,tenKate2010,Moores2012}\revised{. While the responsible UV radiation penetrates no deeper that 500--750 $\mu$m into the subsurface} \citep{Sagan1974,Schuerger2003}\revised{, the topmost centimeters are turned over ("gardened") by random impacts of micrometeorites on geologically short timescales, subjecting organics to photodissociation.} \revised{For the Moon }\citep{Speyerer2016}\revised{estimate that the topmost 2~cm are turned over on a timescale of $\sim$80\,kyr; on Mars, aeolian processes may further accelerate this process. At depths down to tens of centimeters, cosmic-ray bombardment destroys organics on Myr timescales} \citep{Pavlov2012,Pavlov2014}\revised{. Clearly, organics found on the surface and in the immediate subsurface available to scooping or surficial drilling cannot be primordial.}

\revised{Recently, remarkable progress has been made in the search for organic compounds on Mars by the Sample Analysis at Mars instrument suite} \citep[\revised{SAM,}][]{Mahaffy2012}\revised{ onboard \rerevised{NASA\textquotesingle s} \emph{Curiosity} rover. Several different samples, collected in Gale Crater, were analyzed for the presence of organic compounds, including a basaltic aeolian deposit (Rocknest), a smectite-rich mudstone (John Klein and Cumberland), and a phyllosilicate-sulfate-bearing rock (Confidence Hills).} 
\revised{Using SAM, chlorinated hydrocarbons were found in all of these scooped and drilled samples, the first evidence of organics detected on the Martian surface} \citep{Leshin2013,Ming2014,Freissinet2015,Freissinet2016} \revised{. The organics detected in the aeolian samples were attributed to terrestrial contamination }\citep{Glavin2013}\revised{ as aeolian deposits are highly exposed to the harsh Martian conditions. The organics in the other samples and specifically the mudstones are suggested to be Martian in origin, as the clays in these deposits are ideally suited for preserving organics} \citep{Freissinet2015}\revised{. Additionally, methane and its variability were detected in the atmosphere }\citep{Formisano2004,Webster2015}\revised{.}

\revised{An open question is now, what is the source of these organics, especially taking into account  how to reconcile the detection of organics with their (geologically) short lifetimes? Exogenous delivery of organics, from geologically recent impacts of comets, asteroids, and/or interplanetary dust particles (IDPs), is one possible solution to that problem.}

\rerevised{Dust infall rates on Mars were estimated by }\citet{Nesvorny2011}, \citet{Borin2017}, and \citet{Crismani2017}. 
\citet{Nesvorny2011} \rerevised{developed a dynamical model for the Solar System meteoroids, and estimate Earth to accrete  $\sim 15,000$ tons/yr of dust within a factor of two. The corresponding rate for Mars is lower by a factor of $\sim 15$ (Nesvorn\'{y}, private communication), i.e., $\sim 1 \times 10^6$ kg/yr.} 
\rerevised{In a different dynamical model, }\citet{Borin2017} %\rerevised{use }\revised{ a dynamical evolution model of dust particles in the inner Solar System,}
%they estimated the IDP flux on Mars to be $2.96\times10^6$~kg/yr. This corresponds to a carbon flux of $0.3\times10^6$~kg/yr.} 
%\rerevised{from which they 
\rerevised{derive a dust infall rate on Mars of $2.96\times10^6$~kg/yr.}
%assuming a carbon content of 10\% by mass} \citep{Flynn1996}.
%\citet{Crismani2017} \revised{derive a global dust input rate on Mars
%based on spectral observations of $\text{Mg}^+$ in the Martian atmosphere using \rerevised{NASA\textquotesingle s} MAVEN spacecraft, together with a model of the ablation process. Under the assumption that the measured dust input is due to IDPs, their result of $2-3$ tonnes of dust per Martian day corresponds to 0.7--$1.0 \times 10^6$ kg/yr. Adopting the same carbon content as assumed by} \citeauthor{Flynn1996}\revised{, this corresponds to 0.07--$0.1 \times 10^6$ kg/yr of carbon. }
\citet{Crismani2017}, \rerevised{on the other hand, analyze spectral observations of $\text{Mg}^+$ in the Martian atmosphere made using NASA\textquotesingle s MAVEN spacecraft. Combining those observations with a model of the ablation process, they obtain an estimate of the dust infall rate independent of dynamical assumptions: $2-3$ tonnes of dust per Martian day or 0.7--$1.0 \times 10^6$ kg/yr.}
\rerevised{In the following, we adopt the range spanned by the results quoted above as the dust infall rate: $0.7-3 \times 10^6$ kg/yr.  Assuming a carbon content of 10\% by mass } \citep{Flynn1996}, \rerevised{this implies an IDP-borne carbon flux of $0.07-0.3 \times 10^6$ kg/yr.}

Comets and asteroids are the other two potential sources of organics. Comets are known to contain substantial amounts of organics, which was shown by measurements on Halley, 67P/Churyumov-Gerasimenko, and by Stardust \revised{on Wild 2} \citep{Goesmann2015,Jessberger1988,Sandford2006}.
Among the various types of meteorites found on Earth, the carbonaceous chondrites are rich in organics \citep{Sephton2002,Sephton2014}; their parent bodies are the C-class asteroids. \citet{Iglesias-Groth2011} have shown that organic molecules in comets and asteroids are able to survive a radiation dose equal to the expected total dose received during the age of the Solar System. 

The goal of this paper is to study the rates at which comets and asteroids impact Mars in the current Solar System, and to derive the corresponding carbon delivery rates. We do this using standard $N$-body codes modeling the motion of asteroids and comets under the gravitational influence of the Sun and the planets while checking for impacts. These models are described in Section\ \ref{method} together with the derived impact rates. In Section\ \ref{organic}, we calculate the corresponding carbon delivery rates on the Martian surface. In Section\ \ref{ejecta}, we estimate the distribution of asteroid-borne organics near impact craters in comparison to the steady and spatially homogeneous IDP influx. The implications of our findings are discussed in Section\ \ref{discussions}. 

\section{Numerical simulations}\label{method}

We perform numerical simulations of the \rerevised{dynamical evolution} of the current Solar System to study \revised{Mars' impact rates} in geologically recent times; any non-gravitational forces are \revised{not included in the simulations}. We model the Solar System as the Sun plus the eight planets Mercury through Neptune. Planets with natural satellites are modeled as a single body at the position of the planet system's barycenter with the total mass of the planet system. We add different sets of passive test particles, representing asteroids and comets. Under our assumptions, test particles do not interact with one another, therefore comets and asteroids can be studied separately. We integrate forward in time over Myr timescales. 
Over these timescales, the current Solar System is close to a steady state, even when accounting for the non-gravitational Yarkovsky effect, so we can assume impact rates to be constant with time.

To model the \rerevised{dynamical evolution}, we use \revised{the $N$-body} integrator RMVS \citep[Regularized Mixed Variable Symplectic;][]{Levison1994}. It models the motion of one gravitationally dominant object (Sun) and $N$ massive objects (in our case: $N=8$) under the influence of their mutual gravity. Test particles move passively under the influence of the combined gravitational potential of the Sun and the planets. Importantly for our purposes, \revised{the adaptive time step of RMVS allows to properly resolve close encounters leading to impacts of test particles on planets.} Specifically, we used the RMVS implementation that is part of David E.\ Kaufmann's {\tt Swifter} package,%
\footnote{\url{http://www.boulder.swri.edu/swifter/}}
a redesigned and improved version of {\tt SWIFT} \citep{Levison2013}. 
From the code output, we extract the times at which test particles are discarded, the reason why they are discarded (e.g., collision with a planet), and the test particle's internal ID number.

The “baseline” time step in our models is set to \revised{1 day in order to resolve close encounters between asteroids and Mercury} \citep{Granvik2016}. 
RMVS uses adaptive time steps, which are shortened whenever a test particle ventures within three Hill radii of a planet (when the gravitational \rerevised{influences} of planet and Sun become comparable) and are shortened further below one Hill radius. While computationally expensive, this allows impacts to be distinguished from close encounters, as is required for our study.  

Test particles are removed from the simulation once they collide with a planet or with the Sun.%
\footnote{\revised{ }For the purpose of impact checks, planet radii are provided directly. Additionally, users can set a minimum heliocentric distance at which test particles are discarded; we set that value to the solar radius of 0.00465~AU.} 
Moreover, test particles are considered ejected from the Solar System and discarded when they exceed a user-provided heliocentric distance. These values are set separately for asteroid and comet simulations (see Subsections~\ref{method_asteroids} and \ref{method_comets}). 

Table\ \ref{table1} shows the adopted masses and radii, which we took from the Astronomical Almanac.%
\footnote{\url{http://asa.usno.navy.mil/static/files/2016/Astronomical_Constants_2016.pdf}}
The Hill sphere radius $r_H$ follows from the planet mass $m$ and the solar mass $M$, together with \revised{the semimajor axis $a$}:
\begin{equation}\label{eq2:1}
r_H = a \sqrt[3]{\frac{m}{3M}}.
\end{equation}

The {\tt SWIFT} code is agnostic of physical units; units for length, duration, and mass must be used consistently such that the gravitational constant $G$ is unity. We employed AU as length unit and year as time unit, so all masses had to be provided as $G$ times mass, in units of $\rm AU^3 / year^2$.
                  
In all simulations, heliocentric coordinates are used. For asteroids and comets, heliocentric positions and velocities are calculated from perturbed orbital elements \footnote{Using the Kepler equations and employing an iteration method to convert mean anomaly into true anomaly as in \url{http://murison.alpheratz.net/dynamics/twobody/KeplerIterations_summary.pdf}.}, 
based on multi-epoch astrometric data, as given in the Minor Planet Center Orbit Database (MPCORB) \footnote{\url{http://www.minorplanetcenter.net/iau/MPCORB.html}} files. 
\revised{For the planets, heliocentric coordinates and velocities were taken from the NASA's Horizons ephemeris service \footnote{\url{http://ssd.jpl.nasa.gov/horizons.cgi}} for the same epoch as the epoch of the orbital elements of the test particles.}

\begin{table}
\begin{tabular}{ l | l | l | l }
\hline
Object & $GM$ ($\rm AU^3/yr^2$) & Hill sphere radius (AU) & Radius (AU) \\
\hline
Sun & 2.95987$\times 10^{-4}$ & - & 0.00465$^*$  \\
Mercury & 4.91379$\times 10^{-11}$ & 0.00148 & 1.63104$\times 10^{-5}$  \\
Venus & 7.24529$\times 10^{-10}$ & 0.00674 & 4.04551$\times 10^{-5}$ \\
Earth & 8.99929$\times 10^{-10}$ & 0.01001 & 4.26343$\times 10^{-5}$  \\
Mars & 9.55197$\times 10^{-11}$ & 0.00724 & 2.27075$\times 10^{-5}$  \\
Jupiter & 2.82606$\times 10^{-7}$ & 0.34697 & 4.778945$\times 10^{-4}$ \\
Saturn & 8.46185$\times 10^{-8}$ & 0.42881 & 4.028667$\times 10^{-4}$  \\
Uranus & 1.29235$\times 10^{-8}$ & 0.46494 & 1.708513$\times 10^{-4}$  \\ 
Neptune & 1.52474$\times 10^{-8}$ & 0.77035 & 1.655504$\times 10^{-4}$ \\
\hline
\end{tabular}
\caption{Mass, radius and Hill sphere radius for the Sun and planets that were used in all our simulations. $^*$We use the radius of the Sun as the critical lower limit on heliocentric distance, see text.}
\label{table1}
\end{table}

\subsection{Asteroids}\label{method_asteroids}

As of \revised{February 2017}, MPCORB contains orbital elements for \revised{730,272} asteroids. \revised{We consider only orbits calculated from multi-epoch data.} This left us with \revised{618,078} asteroids, which we integrated forward in time over 10 Myr. As discussed below, this integration length was found to be long enough to provide a sufficient number of Mars impactors \revised{and to reach a steady-state population}, while being short compared to the timescale over which, e.g., depletion of impactor orbits or non-gravitational effects such as the Yarkovsky effect would significantly alter our results. Asteroids were considered ejected from the Solar System and discarded when they reached a heliocentric distance of 1,000 AU. 

We are primarily interested in C-type asteroids, which are rich in organics. We ran a first integration \revised{for 1~Myr} of all $\sim$140,000 asteroids \revised{for which an albedo value (based on observations using the WISE spacecraft) was given by }\citet{Masiero2011}\revised{; these asteroids have a diameter range of 1-550 km.} As C-type asteroids are associated with a low albedo \revised{(our adopted value is 0.1)}, this would result in a total number of impacting asteroids with a known fraction of C-type asteroids. However, none of those asteroids was found to collide with Mars. It appears that the population of Mars impactors is composed of asteroids that are too small for WISE detections. 

\begin{figure}
\centering
\includegraphics[width=.8\linewidth]{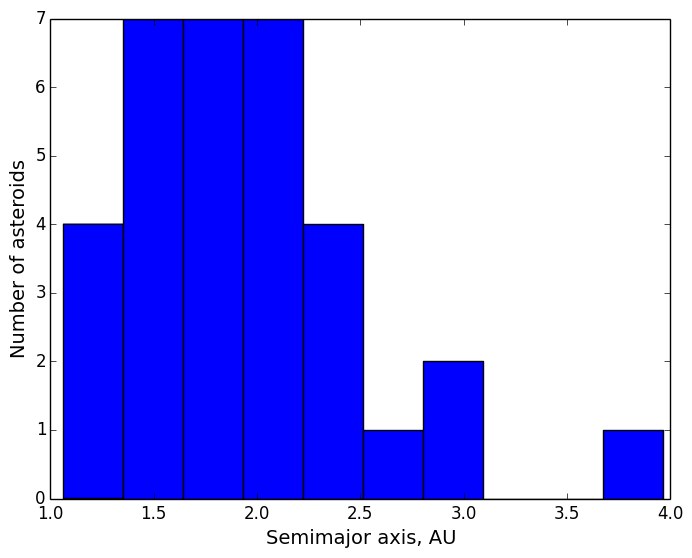}
\caption{Orbital distribution of the \revised{33} Mars impacting asteroids \revised{at the start of the simulations}.}
\label{fig2:1}
\end{figure} 

In a second experiment, we simulated all \revised{618,078} asteroids as described above. During the simulation, \revised{4,609 particles} were ejected from the Solar System. \revised{6,719 particles} collided with the Sun. \revised{1,152 particles} collided with planets. \revised{33} of these impacted with Mars. Other than Mars, five more planets suffered from asteroid impacts: Mercury (\revised{43 particles}), Venus (\revised{463 particles}), Earth (\revised{514 particles}), Jupiter (\revised{97 particles}), and Saturn (\revised{2 particles}).

In Figure\ \ref{fig2:1}, we show the distribution in semimajor axis ($a$) of our \revised{33} Mars impactors \revised{at the beginning of the simulations}. The impactors are seen to originate from semimajor axes of \revised{1.0} to \revised{4.0}~AU with a peak near \revised{1.75}~AU.

As a crosscheck, we have repeated our asteroid simulation with initial conditions taken from MPCORB as of \revised{July 2016} and planet positions for that date (primary simulation: MPCORB from \revised{February 2017}).
During that simulation, \revised{35} asteroids impacted Mars, consistent with our primary result of \revised{33} impactors within the $\sqrt{N}$ Poisson noise (less than 10\%). The two sets of impactors do not overlap\revised{, i.e. the individual asteroids that impact in the first simulation do not impact in the second simulation and vice versa.} Their distribution in semimajor axis, however, is indistinguishable.
We take this to mean that our results are valid in a statistical sense:
while we do not reliably identify individual impacts, we do characterize the population of Mars impactors consistently.

We studied the time evolution of the Mars impact rate during our simulation by analyzing the \revised{cumulative} number of Mars impactors after 1, 2, 3, ..., 10 Myr, respectively.  We expect the number of impactors to be proportional to time. Any significant violation of our assumptions (e.g., depletion of asteroids on impactor orbits or Yarkovsky drift) would result in a deviation from this.  We did find the number of impactors to increase steadily, within Poisson noise, justifying our assumptions. 

\subsection{Comets}\label{method_comets}

As of November 2016, MPCORB contained 879 comets with perturbed orbital elements. We integrated the comets forward in time, together with the Sun and the planets as described above. We chose the integration length to be short enough to not significantly deplete the comet population. After test runs, we converged on \revised{0.1~Myr}. Comets are considered ejected from the Solar System at heliocentric distances beyond 125,000~AU, a much higher value than adopted for asteroids; this is required by the high eccentricity of some comets. 

\begin{table}
\centering
\begin{tabular}{ p{3cm} | p{3cm} | p{3cm} }
\hline
Object & Simulation 1 & Simulation 2 \\
\hline
Total & \revised{346} & \revised{74,835} \\
Sun & \revised{26} & \revised{33,269} \\
Mercury & \revised{0} & \revised{1} \\
Venus & \revised{0} & \revised{7} \\
Earth & \revised{0} & \revised{11} \\
Mars & \revised{0} & \revised{867} \\
Jupiter & \revised{4} & \revised{22,273} \\
Saturn & \revised{0} & \revised{1,357} \\
Uranus & \revised{0} & \revised{5} \\ 
Neptune & \revised{0} & \revised{4} \\
Outside & \revised{316} & \revised{17,041} \\
\hline
\end{tabular}
\caption{Number of discarded comets in our simulations. Simulation~1 was performed for 879 comets \revised{for 1~Myr}. For the second simulation, we cloned our sample \revised{5,000} times by randomizing the angular orbital elements, resulting in \revised{$879*5,000=4,395,000$} synthetic comets; additionally, we increased the radius of Mars by a factor of \revised{twenty. The second simulation was performed for 0.1~Myr The values listed for Simulation~2 are not corrected for population cloning and radius inflation.}}
\label{table2}
\end{table}

As evidenced by a first simulation (see Table\ \ref{table2}, first results column, 'Simulation 1'), \revised{$\sim 39 \%$} of the comets are discarded from the simulation after \revised{1~Myr}, as discussed above.  Importantly, \revised{almost 91\% of the} discarded comets are ejected from the Solar System, making planetary impacts relatively rare. In particular, none of the modeled comets hit Mars.

We attribute this lack of Mars impacts to small-number statistics (note that we integrated $\sim 700$ times more asteroids than comets). We therefore repeated our comet simulation using two methods to improve the statistics (simulation 2). Firstly, we replace each comet with \revised{5,000} synthetic ("cloned") comets with the semimajor axis $a$, eccentricity $e$, and inclination $i$ values of the actual comets, while the remaining, angular, orbital elements are set to random numbers. All resulting impactor fluxes have to be divided by \revised{5,000} to correct for cloning. Secondly, we inflate the radius of Mars by a factor of \revised{20}, so Mars impactor fluxes have to be divided by another factor of \revised{$20^2$} \citep[see][]{Kokubo1996}. \revised{Additionally, we adopted a new simulations' length of 0.1~Myr in order to avoid depletion of the Mars crossing population. As a result of the simulation we identified 867 synthetic comets that collide with Mars. Correcting for cloning and diameter inflation, and dividing by the simulation length, we conclude that there are $867/5,000/400/0.1 \sim 0.00434$ comet impacts on Mars per Myr.} See simulation 2, Table\ \ref{table2} for results of that experiment. Note that the results of these two simulations are mutually consistent after the above-mentioned corrections.

\revised{We estimate the uncertainty in our comet result like we did for asteroids, by repeating our simulation with different initial conditions.  The cross-check simulation used MPCORB and planet positions as of Nov 2017, while the primary simulation was for December 2016.}

\revised{During that simulation, 868 comets impacted Mars, consistent with our primary result of 867 impactors within the $\sqrt{N}$ Poisson noise (less than 10\%). We conclude that our results are valid in a statistical sense.}

\section{Carbon delivery rates}\label{organic}

In the previous section, we estimated the rate at which asteroids and comets impact Mars. In this section, we estimate the amount of carbon delivered by these impacts.

\subsection{Asteroids}\label{organic_asteroids}

Asteroids are commonly divided into a number of taxonomic types with corresponding surface mineralogies. Much progress has been made linking taxonomic types of asteroids to different types of meteorites. Importantly for our project, there is one class of meteorites that contains appreciable amounts of carbon, the carbonaceous chondrite meteorites. Their carbon content is $\sim2\%$ by mass, their parent bodies are the C-class asteroids \citep{Sephton2002,Sephton2014}. \revised{The vast majority of the remaining asteroids are S types, the parent bodies of ordinary chondrite meteorites.  Their carbon content is 0.2\% by weight} \citep{Moore1967}. \revised{As we will see below, they deliver negligible amounts of carbon to Mars. All other taxonomies are too rare to matter and are not considered in this analysis.}

So, in order to estimate the amount of organics delivered by asteroid impacts, we need information on the taxonomy of the impacting population. 
Since our simulation results are only valid in a statistical sense (in particular, we do not attempt to identify individual impactors but only the nature of the impacting population), we adopt a statistical approach.
In doing so, each impactor is assigned a probability $p_\text{C}$ of being C-type.
The expectation value for the amount of carbon delivered by an asteroid equals $p_\text{C}$ times the asteroid mass $m_\text{Asteroid}$ (estimated under the assumption that it is indeed of C-type) times $f_\text{Carbon}$, the measured carbon content of carbonaceous chondrite meteorites:
\begin{equation}
\label{eq3:1}
M_{\rm Organics} = p_\text{C} m_\text{Asteroid} f_\text{Carbon}.
\end{equation}
The carbon delivery rate then follows from the sum over all impactors, divided by the duration of our simulation.

To estimate the mass of an asteroid, we base ourselves on the $H$ magnitude in the standard $HG$ system \citep{Bowell989} of an asteroid as provided by MPCORB; $H$, diameter $D$, and geometric albedo $p_V$ are related to one another via \citep{Pravec2007}:

\begin{equation}
\label{eq3:2}
D = \frac{1329 \rm km}{\sqrt{p_V}} 10^{-H/5} 
\end{equation}
where we assume $p_V = 0.06 \pm 0.01$ \citep[a representative value for C-class asteroids, see][Table 1]{DeMeo2013}. Mass follows from diameter adopting an average C-type mass density $\rho$ of $1.33 \pm 0.58$~g/cm$^3$ \citep[Table 3;][]{Carry2012}:
\begin{equation}
\label{eq3:3}
M = \frac{\pi}{6}  D^3 \rho.
\end{equation}

The most challenging part of our calculations is to estimate $p_\text{C}$, the probability of an asteroid being part of the C class. We estimate $p_\text{C}$ based on the semimajor axis $a$ of an asteroid at the start of the simulation together with the measured fraction of C types (relative to all asteroids) as a function of semimajor axis $a$ across the Main Belt \citep[Figure\ \ref{fig3:1}, based on data from][]{DeMeo2013}. \revised{Note that their results have been debiased against albedo-dependent survey efficiency; otherwise, $p_C$ would be severely underestimated.} We use two methods to calculate this probability. The first method is directly based on \citet{DeMeo2013} but suffers from small-number statistics due to the small number of taxonomic measurements in the population of Mars crossers, which dominates our impactor sample, see Sect.\ref{methodI}. The second method, Sect.\ref{methodII}, is based on published models of the dynamical history of asteroids \citep{Bottke2002,Greenstreet2012} to trace back Mars crossers to 'parent regions' that are much better characterized taxonomically.

The next two subsections give estimates of the C-type fraction, and also quantify their errors, based on the uncertainties in the $H$ magnitude, albedo, mass density, carbon fraction, and the uncertainties in our model of the chaotic dynamics of the Solar System.

\subsubsection{Measured C-type fraction}\label{methodI}

\begin{figure}
\centering
\includegraphics[width=.99\linewidth]{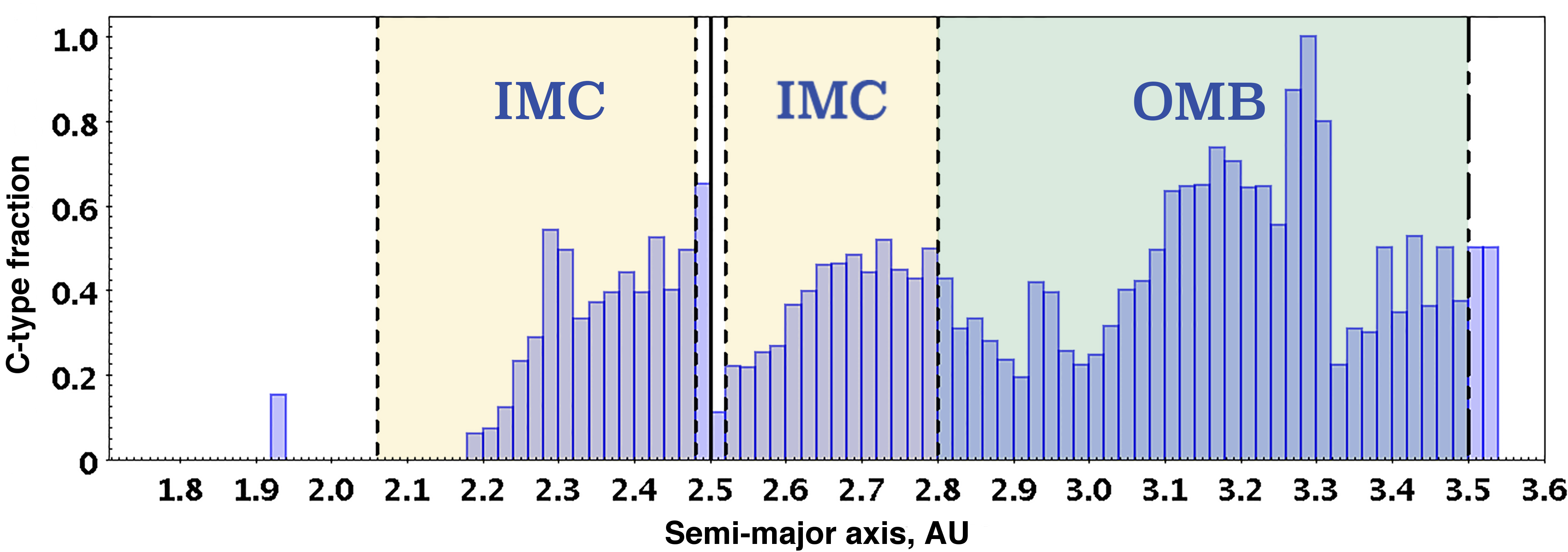}
\caption{Measured fraction of C-type asteroids across the Main Asteroid Belt. See \citet{DeMeo2013} for details; this plot has been generated using the data (kindly provided by Francesca DeMeo) shown in their Figure 9, where we exclude all types but C. Three source regions \citep[following][]{Bottke2002,Greenstreet2012} are indicated on the plot. The yellow area corresponds to intermediate Mars crossers (IMC), green represents the outer main-belt (OMB), while the solid line at $\sim 2.5\ \rm AU$ represents the 3:1 resonance. The $\nu_6$ resonance is shown in Figure\ \ref{fig3:2}. The last source region, Jupiter Family Comets, is not relevant for our study.}
\label{fig3:1}
\end{figure}

To calculate the carbon content of the impacting asteroids, we compare the semimajor axes $a$ of the \revised{33} Mars impactors (see Figure\ \ref{fig2:1}) with the measured C-type fraction as a function of semimajor axis $a$ (see Figure\ \ref{fig3:1}).

Unfortunately, the measured C-type fraction is poorly constrained (see Figure~\ref{fig3:1}) for $a\leq 2$ AU where many impactors reside. For asteroids with $a<2$ AU, we assume a C-type fraction of 0.1538, taken from Figure\ \ref{fig3:1} for $a=1.92$ AU. For asteroids with $2$ AU $< a < 2.18$ AU, we adopt a C-type fraction of 0.0612, the \citeauthor{DeMeo2013} value for $a=2.18$ AU. For $a>2.18$ AU, we use C-type fractions specified for each 0.02 AU bin in \cite{DeMeo2013}, shown in Figure\ \ref{fig3:1}. 

Combining all of the above using Equation~\ref{eq3:1}, we arrive at a rate of carbon delivery to Mars due to asteroid impacts of \revised{$0.121 \pm 0.041 \times 10^6$ kg/yr,} averaged over 10 Myr. The uncertainty is calculated combining the uncertainties of $H$ magnitude, albedo, \revised{and} mass density. 

\subsubsection{Measured C-type fraction plus source regions}\label{methodII}

To overcome the small-number statistics plaguing the analysis above, we have used published models of the dynamical origin of near-Earth asteroids as \revised{described} by \citet{Bottke2002}. There are five major 'source regions' from which asteroids diffuse due to resonances with the planets: the $\nu_6$ resonance, the 3:1 resonance with Jupiter($\sim 2.5$ AU), the intermediate Mars crossers (IMC) population ($2.06 < a < 2.48$ AU or $2.52 < a < 2.80$ AU), the outer main-belt (OMB) asteroids ($a > 2.8$ AU), and the Jupiter family comets. We base our analysis on the model by \citet{Greenstreet2012}, who essentially redid the \citet{Bottke2002} analysis using a higher time resolution, including the gravitational effects of Mercury, and using a much larger number of test particles. Their data products%
\footnote{Available from \url{http://www.sarahgreenstreet.com/neo-model/}}
can be used to calculate the probability of an asteroid originating from the five source regions as a function of its position in a finely sampled grid in semimajor axis $a$, eccentricity $e$, and inclination $i$.

The original models by \citet{Bottke2002} and \citet{Greenstreet2012} only apply to near-Earth asteroids out to a perihelion distance of $q\leq 1.3$ AU. We re-binned and re-normalized their model with a higher cut-off perihelion distance of $q\leq 2$ AU. We note that their model was run otherwise unchanged. In particular, no attempt was made at identifying additional regions in the Main Belt that may be productive source regions for Mars crossers (but not for Earth crossers). This may limit the accuracy of our results.

All \revised{33} Mars impactors have a perihelion distance of $q\leq 2$ AU. Below, we describe how their carbon content was estimated using the source regions.

For a $q\leq 2$ AU asteroid, the modified \citet{Greenstreet2012} model provides five probabilities (adding up to 100\%) of the object originating from one of five source regions. 
For four of these regions, we obtain a straightforward C-type fraction from Figure\ \ref{fig3:1}:
\begin{itemize}
\item 3:1 resonance: $a \sim 2.5$ AU, C-type fraction 0.1111
\item Intermediate Mars crossers: $2.06 < a < 2.48$ AU or $2.52 < a < 2.8$ AU, C-type fraction 0.3496 (averaged over relevant $a$ range)
\item Outer main-belt: $ 2.8$ AU $< a < 3.5$ AU, C-type fraction 0.4663
\item Jupiter family comets: $a > 4.2$ AU , C-type fraction 1.0000. Note that all Mars impactors had 0.00\ \%  probability of originating from JFCs.  
\end{itemize} 

\begin{figure}
\centering
\includegraphics[width=.76\linewidth]{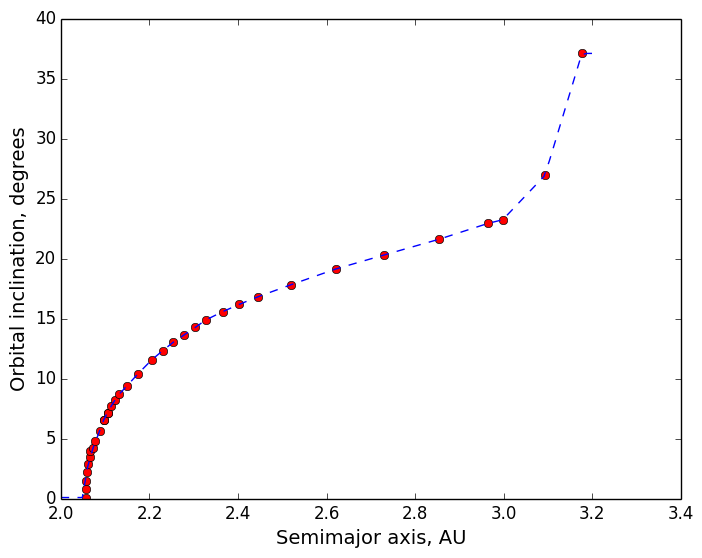}
\caption{$\nu_6$ resonance in a,i space (Kevin J. Walsh, private communication, approximating Fig.~4 in \citet{Milani1990}).}
\label{fig3:2}
\end{figure}

The fifth "source region," the secular $\nu_6$ resonance, spans a wide range of semimajor axes, starting just longward of 2.0 AU. Importantly, the corresponding $a$ increases with orbital inclination $i$ (see Figure\ \ref{fig3:2}). Also, $i$ is approximately preserved under scattering in the $\nu_6$ resonance. To assign a C-type fraction to the probability of each of our impactors originating at $\nu_6$, we therefore assume its current inclination $i$ was preserved upon migration from the Main Belt, and read off $a$ of its $\nu_6$-origin from Figure\ \ref{fig3:2}. The corresponding C-type fraction then follows from Figure\ \ref{fig3:1} and varies between 0.0306 and 0.7049.

\citet{Granvik2017} studied the orbital inclination of asteroids before and after scattering out of the asteroid belt due to resonances including the $\nu_6$ secular resonance. They show that scattering at the $\nu_6$ resonance tends to increase the inclination. We neglect this inclination change. \citeauthor{Granvik2017} \revised{show that, for the subset of asteroids which escape through the $\nu_6$ resonance, biggest inclination shift occurs for asteroids with $i<~6$ degrees, that corresponds to a range of semi-major axes (2.056-2.088~A.U.) for which the probability of being C-type equals zero.} Assuming a constant inclination will therefore cause us to underestimate the carbon delivery rate due to asteroid impacts.  While the size of that effect is difficult to estimate quantitatively based on the \citeauthor{Granvik2017}\ histogram, we take it to be small compared to our uncertainties.

To calculate the total carbon mass delivered by the \revised{33} Mars impactors with $q\leq 2$ AU, we modify Equation~\ref{eq3:1}:

\begin{equation}
\label{eq3:4}
M_{\rm Organics} = \sum_{\rm Impactors} \sum_{\rm SourceRegions} p_{\rm SourceRegion} p_\text{C|SourceRegion} m_\text{Asteroid} f_\text{Carbon},
\end{equation}
where $p_{\rm SourceRegion}$ is the probability that an asteroid comes from a specific source region, $p_\text{C|SourceRegion}$ is its probability of being C-type under the assumption that it does originate from that source region, $m_\text{Asteroid}$ is the asteroid mass as in Equation~\ref{eq3:3} and the paragraphs above it, and $f_\text{Carbon}$ is the carbon content of a carbonaceous chondrite ($\sim 2\%$).

Using Equation~\ref{eq3:4}, we arrive at a total carbon flux to Mars due to asteroid impacts of \revised{$0.099 \pm 0.041 \times 10^6$ kg/yr.} The uncertainty is calculated combining the uncertainties of the $H$ magnitude, albedo, mass density, and source region probabilities.	

Our results derived using the two methods described in Subsections~\ref{methodI}-\ref{methodII} are mutually consistent.
\revised{The mean value of the results obtained with the two methods for the carbon flux at Mars due to asteroid impacts is $0.11 \pm 0.057 \times 10^6$ kg/yr}. 
Using the same methods, we also analyzed the results of the crosscheck simulation, based on the older MPCORB version, \revised{resulting in a mean value for the carbon delivery rate of $0.023 \pm 0.007 \times 10^6$ kg/yr, lower than our primary simulation by a factor of nearly five. We attribute this discrepancy to the different size-distribution of the Mars impactors: the primary simulation contained several relatively big impactors of 2-5~km in diameter, while impactors in the crosscheck simulation were smaller. The largest few impactors dominate the mass budget, leading to large Poisson noise.  In summary, our final adopted asteroid delivery rate is $0.05 \times 10^6$ kg/yr within a factor of $\sim$two.} 

\revised{We also calculated the carbon flux due to impacts of S-type asteroids.  To this end, we used the results of the nominal asteroid simulation together with the measured distribution of S-types within the main belt (analogous to the C-type distribution discussed above).  Assuming an S-type albedo of $p_V=0.23\pm0.02$ } \citep{DeMeo2013} \revised{and a mass density of $\rho = 2.72\pm0.54$ g/cm${}^3$} \citep{Carry2012}, \revised{we obtain a carbon delivery rate of $0.004\times 10^6$ kg/year, negligible compared to C types as expected.}

\subsection{Comets}\label{organic_comets}

As derived in Section\ \ref{method_comets}, the impactor flux on Mars due to comets is \revised{$\sim$ 0.00434} comets per Myr. Assuming, following \citet{Swamy2010}, a typical comet mass of $3 \times 10^{13}$ kg and a carbon content of 10\%, this implies a carbon delivery rate to Mars due to comet impacts of \revised{$0.013\times 10^6$ kg/yr}, smaller than the rate due to IDPs \revised{and asteroids}.

We note that our method of using an average mass per comet is less rigorous than our approach for asteroids. However, seeing the cloning method employed, we do not feel that more sophistication is warranted. 

\subsection{Mass contribution of small undiscovered asteroids and comets}\label{MFD}

The calculations above are based on catalogs of \emph{known} asteroids and comets. These catalogs are known to be incomplete, however, especially for small objects that tend to be too faint for observation. In this section, we estimate the contribution of the undiscovered population to our results.

An extensive amount of work has been devoted to studies of the size-frequency distribution (SFD) of asteroids. The number density $n(D)$ of asteroids with diameter $D$ is typically described as a power law
\begin{equation}
n(D) \propto D^{\alpha}
\end{equation}
with exponent ("slope") $\alpha$. Note that we are discussing the differential SFD; the cumulative SFD (describing asteroids larger than some $D_0$) is also often discussed, \revised{whose} exponent equals $\alpha+1$.
Various estimates for the exponent $\alpha$ exist in the literature for different asteroid populations and size ranges; in all estimates, $\alpha$ scatters around $\alpha=-3.5$ \citep[see, e.g.][]{Bottke2005}. As we will see below, our results are not sensitive to reasonable changes in $\alpha$. The total mass of a population of asteroids with a maximum diameter of $D_\text{max}$ equals
\begin{equation}\label{eq3:5}
\int_0^{D_\text{max}} n(D) M(D) dD \propto D^{4+\alpha} \Big|_{0}^{D_\text{max}}
\end{equation}
where $M(D)\propto D^3$ is the mass of an asteroid of diameter $D$. Importantly, the integral converges as $D\to 0$ as long as $\alpha \geq -4$.

Denoting as $D_\text{min}$ the diameter down to which observed SFDs are observationally complete to a reasonable level, our results can be corrected for undiscovered asteroids by multiplying them by a factor
\begin{equation}
\frac{\int_0^{D_\text{max}} n(D) M(D) dD }{\int_{D_\text{min}}^{D_\text{max}} n(D) M(D) dD } = \frac{D_\text{max}^{4+\alpha}}{D_\text{max}^{4+\alpha} - D_\text{min}^{4+\alpha}} = \frac{1}{1-\left(D_{\rm min}/D_{\rm max}\right)^{4+\alpha}}.
\end{equation}
For the inner belt, a $D_\text{min}$ of 1--2~km is reasonable; for $D_\text{max}$ we use \revised{Ceres'} diameter of $\sim$1,000 km. For $\alpha\sim -3.5$, the resulting correction is $\sim$4\%, completely negligible given our uncertainties.
\revised{The biggest contribution to the impact flux will always be provided by the few largest asteroids} \citep{Turrini2014}.
While the SFD of comets is less well constrained, it is reasonable to assume that it is not too different from that of asteroids, leading us to neglect the influence of undiscovered comets, too.

\section{Carbon content of the ejecta blanket}\label{ejecta}
One important difference between material delivered through IDPs and through comets and asteroids is its spatial and temporal distribution. The IDP flux is homogeneous over the surface of Mars and constant over time scales that are long compared to its orbital period. \revised{One the other hand, impacts by comets and asteroids} happen rarely (\revised{$\sim33$} large asteroid impacts over 10 Myr and even  less for comets) and deposit material locally, around the resulting impact crater.

In this section, we study the spatial distribution of comet and asteroid material and compare it to the "IDP background". 
To this end, we perform a case study assuming the most likely type of impactor, an asteroid with diameter around 1~km.
Following, e.g., \citet[][Table~1]{Holsapple2007}, 
the size of the resulting crater can be estimated
in two end-member cases where crater formation is dominated by gravity and cohesion ("strength"), respectively.
Roughly speaking, the gravity regime applies to impact events that form large craters, and the strength regime to impact events that form small craters; intermediate cases will be more complex. 
In the gravity and strength regime, respectively, the radius $R$ of the final crater equals:
\begin{align}
\label{eq4:1}
\frac{R}{a} &= 1.03 \left(\frac{ga}{U^2}\right)^{-0.170} \left(\frac{\delta}{\rho}\right)^{0.332} \\
\label{eq4:2}
\frac{R}{a} &= 1.03 \left(\frac{\bar{Y}}{\rho U^2}\right)^{-0.205}\left(\frac{\delta}{\rho}\right)^{0.40}
\end{align}
where 
$a$ is the impactor radius (500 m in our case), $g$ is the gravitational acceleration on the surface of Mars (adopted value: 3.711 m/s$^2$), $U$ is the normal component of the impactor velocity (assumed value: 7.1 km/s \revised{for the assumed impact angle of 30$^{\circ}$}), $\delta$ is the mass density of the impactor (1.33 g/cm$^3$, the density of C-type asteroids), $\rho$ is the density of the target material (assumed: 3.93 g/cm$^3$). $\bar{Y}$ is the strength of the target material \citep[assumed values, following][: 50\,Pa, 3,000\,Pa, 12,000\,Pa, 65,000\,Pa, spanning a wide range bracketing plausible values for the surface of Mars]{Holsapple2007}.

Per the above, the crater radius $R$ becomes \revised{2.039\,km} for the gravity regime, while in the strength regime we obtain crater radii of \revised{7.520\,km, 3.248\,km, 2.445\,km, and 1.729\,km} for the four strength values listed above.

\begin{figure}
\centering
\includegraphics[width=1.\linewidth]{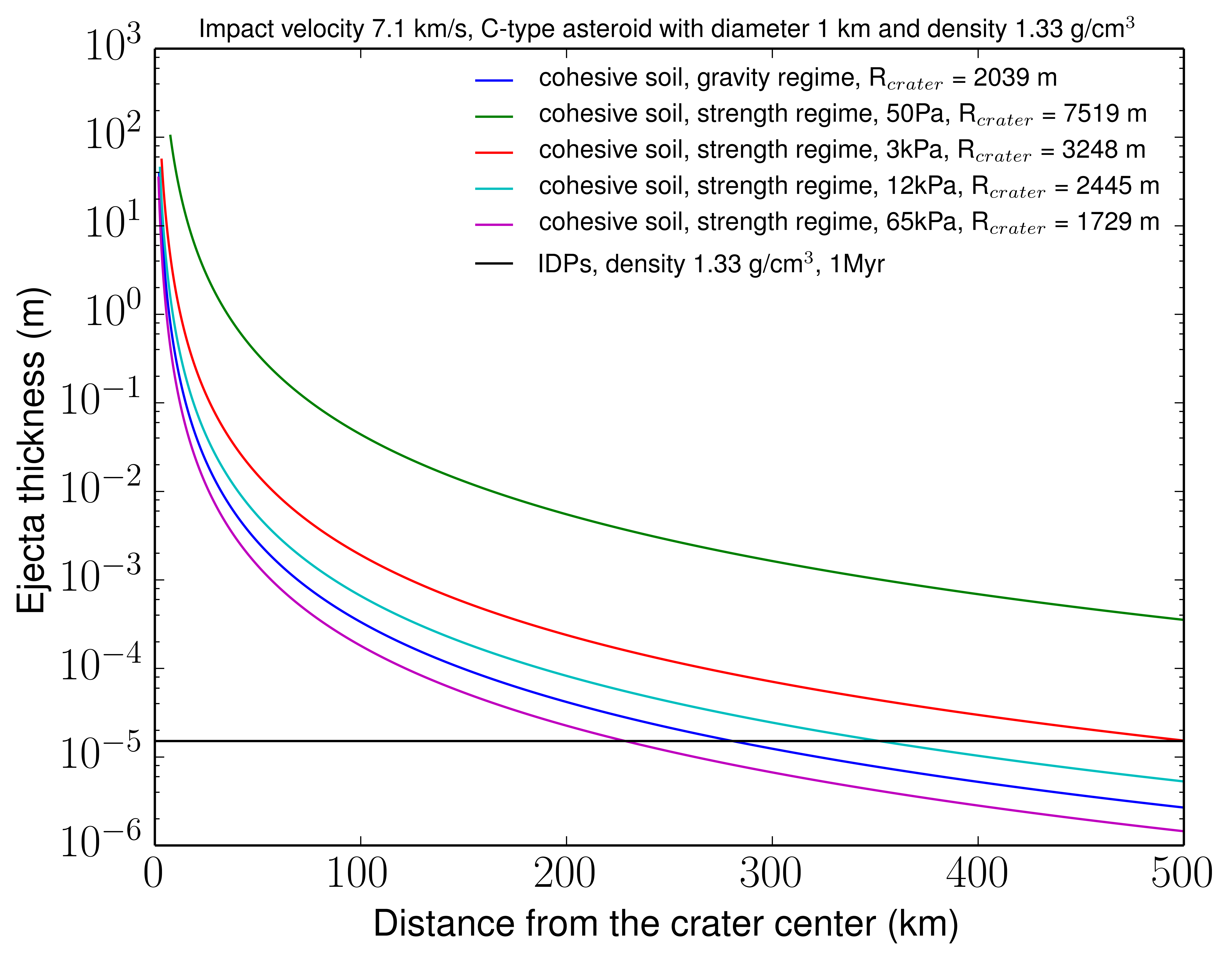}
\caption{\revised{Ejecta thickness as a function of  distance from the crater center. Colored curves correspond to different types of Martian soil, see text. The black horizontal line corresponds to dust layer formed by IDPs over 1~Myr.}}
\label{fig4:1}
\end{figure}

These crater radii can be used to estimate the ejecta thickness as a function of distance from the crater center, $r$, using Equation~1 of the on-line material in \citet{Marchi2012}:
\begin{equation}
\label{eq4:3}
t(r) = 0.14 {r_c}^{0.74} \left(\frac{r}{R}\right)^{-3} \frac{r/R_M}{\sin(r/R_M)},
\end{equation}
where $t$ is the ejecta thickness, $R$ is the crater radius as estimated above ($r>R$), and $R_M$ is the radius of Mars. Note that \citet{Marchi2012} is concerned with craters on the Moon, so in their work, $R_M$ denotes the lunar radius. For the crater radii derived above, the ejecta blanket can be hundreds of meter thick near the crater rim and decreases with increasing distance (see Fig.~\ref{fig4:1}). At 500~km from the crater, the ejecta thickness is at the micron level.

The ejected material is a mix of Martian soil and asteroid material. We are interested in its carbon content deriving from the asteroid impactor (for simplicity, Mars material is assumed to be completely depleted in carbon). The mass of asteroid-borne organics per surface area, $o(r)$, is proportional to ejecta thickness; its integral over the area outside the crater must equal the total carbon content of the asteroid, $m_\text{O}$:
\begin{equation}
2\pi\int_R^\infty r o(r) \text{d}r = m_\text{O}.
\end{equation}
In doing so, we tacitly assume that no ejected material can escape to space, but everything is retained by Martian gravity. This would be an issue when studying impact delivery of material on smaller objects such as Vesta \citep[see, e.g.,][]{Turrini2014_model} but should be non-critical on an object as massive as Mars. \revised{Using Equation~8 from} \citet{Svetsov2011} \revised{we see that for low velocities (below 15~km/s) up to 23 \% of the impactor will escape. We caution that ejecta will interact with the atmosphere, although studying that would be beyond our scope.  Atmospheric friction decelerates the smallest ejecta particles, causing them to land closer to the crater.  Aeolian processes may disturb the post-deposition ejecta distribution.} 

\begin{figure}
\centering
\includegraphics[width=1.\linewidth]{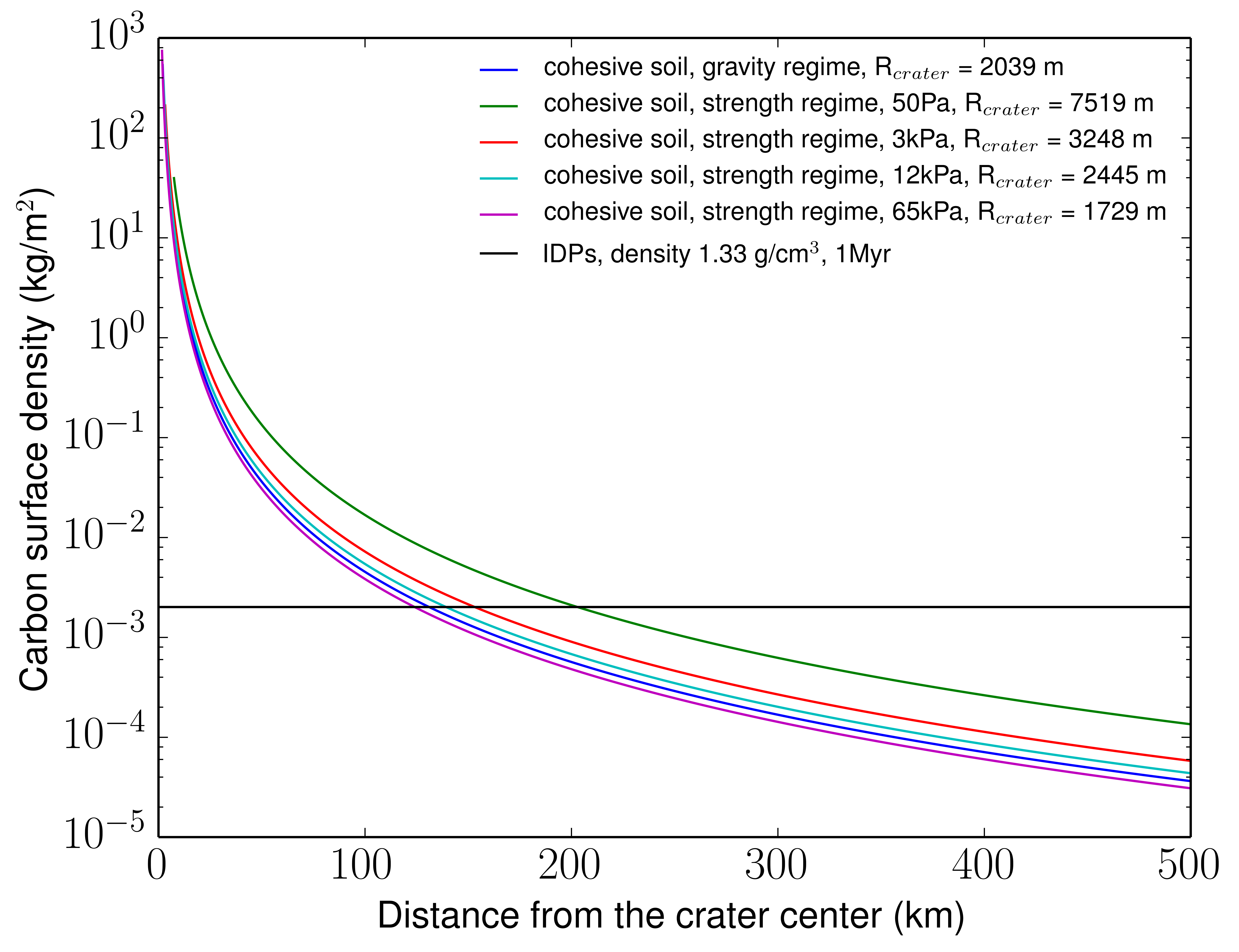}
\caption{Carbon surface density as a function of  distance from the crater center. Colored curves correspond to different types of Martian soil, see text. The black horizontal line corresponds to carbon delivered by IDPs over 1~Myr.}
\label{fig4:2}
\end{figure}

The colored curves in Figure~\ref{fig4:2} show $o(r)$ for the crater radii mentioned above, corresponding to different soil properties. They are very similar to one another; the details of the cratering mechanism, including the assumed strength values where applicable, are of secondary importance for our purposes.
The black horizontal line represents the carbon delivered by the "IDP background" over 1~Myr, a conservative estimate of the lifetime of organics near the Martian surface. Clearly, for these time scales, asteroid-borne organics dominate over IDP-borne organics for distances up to 100-\revised{200} km. This is also applicable for comet-borne organics.  

\rerevised{To be conservative, for the "IDP background" we have used the upper limit of the previously derived IDP-borne carbon flux ($0.07 - 0.3 \times 10^6$ kg/yr); using the lower value would slightly increase the distance out to which large impactors dominate over IDPs.}
%To be conservative, we used the \revised{ }\citeauthor{Borin2017}\revised{ } value for the IDP-borne carbon flux; using the lower \citeauthor{Crismani2017} \rerevised{or }\citeauthor{Nesvorny2011} value would \revised{slightly} increase the distance out to which large impactors dominate over IDPs.

\revised{As a first-order estimate, we take the typical "area of influence" of an asteroid impact to be a circle of r$=100$~km.  Given 33 asteroid impacts, asteroid-borne organics will dominate over IDP-borne organics on $\sim 1$\% of the total surface area of Mars.}

\section{Discussion}\label{discussions}

Our main result is the delivery rate of organics to Mars due to impacts by asteroids and comets: \revised{$\sim$0.05$\times10^6$ kg/yr} and \revised{$\sim$0.013$\times10^6$ kg/yr}, respectively. This is to be compared against delivery by IDPs.

\rerevised{A dynamical model for the solar system meteoroids from} \citet{Nesvorny2011} \rerevised{predicts the total mass accreted by Mars to be $\sim 1 \times 10^6$ kg/yr.} \citet{Borin2017} \revised{ estimate the total dust flux due to be $2.96 \times 10^6$ kg/yr based on observations of IDP influx on Earth and dynamical models.} Also, \citet{Crismani2017} derive the IDP infall rate based on MAVEN-based observations of \revised{ $\text{Mg}^+$} in the atmosphere of Mars, which they take to be caused by incoming dust and convert to a dust flux using an ablation model. \revised{Assuming the carbon content to be 10\%} \citep{Flynn1996}, \citeauthor{Nesvorny2011}, \citeauthor{Borin2017} and \citeauthor{Crismani2017} \rerevised{results imply an IDP-borne carbon flux of $0.07 - 0.3 \times 10^6$ kg/yr.} 

Our result for organics delivered by comet impacts is \rerevised{at least one order of magnitude less than the asteroid value given the error bars. Comets deliver between $\sim$4 \% and $\sim$19 \% of the IDP-born organic flux as estimated by }\citet{Nesvorny2011}\rerevised{, }\citet{Borin2017}\rerevised{ and }\citet{Crismani2017}\rerevised{. On the other hand, the asteroid contribution is $\sim$17 - 71 \% of the IDP-borne flux.} %estimated by }\citet{Nesvorny2011}\rerevised{, }\citet{Borin2017}\rerevised{ and }\citet{Crismani2017}\rerevised{.}

Our asteroid cross-check simulation, performed identically to the primary one except for a different start date, resulted in an impactor population that does not overlap with the 'primary' impactor population but has identical ensemble properties; in particular, the amount of carbon delivered by the two populations is indistinguishable. 
We conclude that our results are correct in a statistical sense. We do \emph{not} reliably identify individual impactors. Such a study, if at all possible over 10 Myr, would require significantly more effort and is beyond our scope.

In an independent study, \citet{Granvik2017} integrated the orbital evolution of a subset of the asteroid main belt (their "set C") containing 78,335 asteroids for 100~Myr including an approximative treatment of the non-gravitational Yarkovsky and YORP effects.
In doing so, they cloned the Hungaria and Phocaea populations sevenfold and threefold, respectively, increasing the number of simulated asteroids by some 18\%, to 92,449.
They report 114 Mars impacts, the delivered organics are not discussed.  
Rescaling their result to match our sample size, this corresponds to $~76.2$ asteroid impacts / 10 Myr (where we made no attempt at correcting for cloning), somewhat larger than our result of \revised{33 impacts} / 10 Myr. 
Qualitatively, this difference is consistent with the expectation that the Yarkovsky and YORP effects, which we neglect, replenish the impactor population.
Quantitatively, this difference does not change our final result for carbon delivery rates beyond our stated error bars.
Our study of the time evolution of the asteroid impact rate (see Sect.\ \ref{method_asteroids})
shows that the asteroid impact rate stays roughly constant (within Poisson noise) over the course of the simulated 10 Myr, providing independent support for our choice to neglect Yarkovsky and YORP.

IDPs are decelerated by Mars's tenuous atmosphere much more efficiently than asteroids are. As detailed by \citet{Flynn1996}, a non-negligible fraction of IDPs reach the surface at low enough temperatures for any present organics to reach Mars intact. 
Asteroids or comets, on the other hand, will hit the surface of Mars at escape velocity or faster, with no efficient aerobraking except for very oblique impacts \citep{Blank2008}. \revised{Depending on impact angle and velocity, some fraction of the impactor mass will evaporate upon crater formation, destroying any organic molecules present. However, for low impact velocities }\citep[\revised{below 10km/s, }][]{Svetsov2015,Turrini2016}\revised{ up to 40\% of the projectile mass remains in the crater and nearby; 
impactor's material will be heated only to temperatures of 1000\,K leaving most carbon unaltered.}

Furthermore, impactor material will be buried at some depth, providing the delivered organics some temporary shelter against photodissociation until they are "dug up" due to impact gardening \citep[on kyr timescales, ][]{Speyerer2016}.
It is beyond our current scope to discuss these effects in detail. \revised{We wish} to emphasize that further studies of the role of asteroids and comets in the delivery of organics to Mars are warranted.

As shown in Section\ \ref{MFD}, the mass flux due to asteroids and comets is dominated by the largest impactors\revised{: t}heir diameters are in the km range. Such impacts are too small to cause global ejecta blankets, 
thus asteroid and comet-borne organics will remain localized on Mars's surface. 

\revised{The samples analyzed by SAM were collected at three different locations in Gale Crater, an impact crater formed around 3.6 billion years ago. The crater itself shows evidence of processing after the impact, including aqueous alteration, as suggested by the presence of hydrated sulphates and smectite clay minerals. The younger layers higher on the mound are covered by a mantle of dust }\citep{Wray2013}\revised{. Depending on the (unknown) taxonomy of the impactor creating Gale Crater, the organics detected by SAM could be asteroidal/cometary in origin, provided they were buried sufficiently deeply at deposition, and that they moved closer to the surface through later geology.  Other origins, e.g., through micro-meteorites or IDPs are not excluded, however.}

In the vicinity of impacts, comet and asteroid-borne organics will dominate over IDP-borne organics, which are distributed very evenly across Mars's surface. In Section\ \ref{ejecta}, we show that the carbon surface density of the ejecta blanket, created \revised{by} the impact of 1~km size asteroid, can reach 1,000~kg/m$^2$ close to the edge of the crater, much larger than the carbon layer due to IDPs of $\sim$10$^{-3}$~kg/m$^2$. Comet and asteroid-borne organics dominate over their IDP-borne counterparts at distances up to 100--200~km from the crater center. This could become important in the analysis of \emph{in-situ} observations, which may uncover locally elevated organics content. Our results suggest that this could be caused by a geologically recent impact of a C-type asteroid or by a comet impact, at a distance of up to some \revised{150~km.}

\section{Conclusions}\label{conclusions}

We conclude that comet \revised{impacts deliver amounts of organics between $\sim 4$ \% and $\sim 19$ \% of the IDP-borne flux, depending on how the latter is estimated. The asteroids deliver between $\sim 17$ \% and $\sim 71$ \%.} Comet and asteroid-borne organics are mostly deposited locally, around impact locations. In those areas, up to \revised{150~km} from the crater center, comet/asteroid-borne organics will dominate over IDP-borne organics. These finding could prove important for the analysis of \emph{in-situ} measurements.

\section{Acknowledgments}\label{aknowledgments}
We are thankful to \emph{Francesca DeMeo} for providing us the taxonomic distribution of asteroids and for useful suggestions, \emph{Alessandro Morbidelli} for interesting discussions and suggestions, \emph{Oleksandra Ivanova} for input on comets, \emph{Kevin Walsh} for providing a table for Figure~\ref{fig3:2}, \emph{Simone Marchi} for the crater ejecta input and \rerevised{\emph{David Nesvorn\'{y}} for help on the IDP influx.}   

This research has made use of data and/or services provided by the International Astronomical Union's Minor Planet Center.

We would like to thank the Center for Information Technology of the University of Groningen for their support 
and for providing access to the Peregrine high performance computing cluster.

\rererevised{The authors thank the two anonymous reviewers for their thoughtful comments, which significantly improved the manuscript.}

\section*{References}

\bibliography{mybibfile}

\begin{thebibliography}{51}
\expandafter\ifx\csname natexlab\endcsname\relax\def\natexlab#1{#1}\fi
\expandafter\ifx\csname url\endcsname\relax
  \def\url#1{\texttt{#1}}\fi
\expandafter\ifx\csname urlprefix\endcsname\relax\def\urlprefix{URL }\fi

\bibitem[{{Blank} et~al.(2008){Blank}, {Liu}, {Lomov}, and
  {Antoun}}]{Blank2008}
{Blank}, J.~G., {Liu}, B.~T., {Lomov}, I.~N., {Antoun}, T.~H., Mar. 2008.
  {Modeling Comet-Earth Collisions to Assess Survivability of Organic Materials
  During Impacts}. In: Lunar and Planetary Science Conference. Vol.~39 of Lunar
  and Planetary Science Conference. p. 2237.

\bibitem[{{Borin} et~al.(2017){Borin}, {Cremonese}, {Marzari}, and
  {Lucchetti}}]{Borin2017}
{Borin}, P., {Cremonese}, G., {Marzari}, F., {Lucchetti}, A., 2017. {Asteroidal
  and cometary dust flux in the inner solar system}. \aap.

\bibitem[{{Bottke} et~al.(2005){Bottke}, {Durda}, {Nesvorn{\'y}}, {Jedicke},
  {Morbidelli}, {Vokrouhlick{\'y}}, and {Levison}}]{Bottke2005}
{Bottke}, W.~F., {Durda}, D.~D., {Nesvorn{\'y}}, D., {Jedicke}, R.,
  {Morbidelli}, A., {Vokrouhlick{\'y}}, D., {Levison}, H., May 2005. {The
  fossilized size distribution of the main asteroid belt}. \icarus 175,
  111--140.

\bibitem[{{Bottke} et~al.(2002){Bottke}, {Morbidelli}, {Jedicke}, {Petit},
  {Levison}, {Michel}, and {Metcalfe}}]{Bottke2002}
{Bottke}, W.~F., {Morbidelli}, A., {Jedicke}, R., {Petit}, J.-M., {Levison},
  H.~F., {Michel}, P., {Metcalfe}, T.~S., Apr. 2002. {Debiased Orbital and
  Absolute Magnitude Distribution of the Near-Earth Objects}. \icarus 156,
  399--433.

\bibitem[{{Bowell} et~al.(1989){Bowell}, {Hapke}, {Domingue}, {Lumme},
  {Peltoniemi}, and {Harris}}]{Bowell989}
{Bowell}, E., {Hapke}, B., {Domingue}, D., {Lumme}, K., {Peltoniemi}, J.,
  {Harris}, A.~W., 1989. {Application of photometric models to asteroids}. In:
  {Binzel}, R.~P., {Gehrels}, T., {Matthews}, M.~S. (Eds.), Asteroids II. pp.
  524--556.

\bibitem[{{Carry}(2012)}]{Carry2012}
{Carry}, B., Dec. 2012. {Density of asteroids}. \planss 73, 98--118.

\bibitem[{{Crismani} et~al.(2017){Crismani}, {Schneider}, {Plane}, {Evans},
  {Jain}, {Chaffin}, {Carrillo-Sanchez}, {Deighan}, {Yelle}, {Stewart},
  {McClintock}, {Clarke}, {Holsclaw}, {Stiepen}, {Montmessin}, and
  {Jakosky}}]{Crismani2017}
{Crismani}, M.~M.~J., {Schneider}, N.~M., {Plane}, J.~M.~C., {Evans}, J.~S.,
  {Jain}, S.~K., {Chaffin}, M.~S., {Carrillo-Sanchez}, J.~D., {Deighan}, J.~I.,
  {Yelle}, R.~V., {Stewart}, A.~I.~F., {McClintock}, W., {Clarke}, J.,
  {Holsclaw}, G.~M., {Stiepen}, A., {Montmessin}, F., {Jakosky}, B.~M., May
  2017. {Detection of a persistent meteoritic metal layer in the Martian
  atmosphere}. Nature Geoscience 10, 401--404.

\bibitem[{{DeMeo} and {Carry}(2013)}]{DeMeo2013}
{DeMeo}, F.~E., {Carry}, B., Sep. 2013. {The taxonomic distribution of
  asteroids from multi-filter all-sky photometric surveys}. \icarus 226,
  723--741.

\bibitem[{{Flynn}(1996)}]{Flynn1996}
{Flynn}, G.~J., 1996. {The Delivery of Organic Matter from Asteroids and Comets
  to the Early Surface of Mars}. Earth Moon and Planets 72, 469--474.

\bibitem[{{Formisano} et~al.(2004){Formisano}, {Atreya}, {Encrenaz},
  {Ignatiev}, and {Giuranna}}]{Formisano2004}
{Formisano}, V., {Atreya}, S., {Encrenaz}, T., {Ignatiev}, N., {Giuranna}, M.,
  Dec. 2004. {Detection of Methane in the Atmosphere of Mars}. Science 306,
  1758--1761.

\bibitem[{{Freissinet} et~al.(2016){Freissinet}, {Glavin}, {Buch}, {Szopa},
  {Summons}, {Eigenbrode}, {Archer}, {Brinckerhoff}, {Brunner}, {Cabane},
  {Franz}, {Kashyap}, {Malespin}, {Martin}, {Millan}, {Miller},
  {Navarro-Gonzalez}, {Prats}, {Steele}, {Teinturier}, and
  {Mahaffy}}]{Freissinet2016}
{Freissinet}, C., {Glavin}, D.~P., {Buch}, A., {Szopa}, C., {Summons}, R.~E.,
  {Eigenbrode}, J.~L., {Archer}, P.~D., {Brinckerhoff}, W.~B., {Brunner},
  A.~E., {Cabane}, M., {Franz}, H.~B., {Kashyap}, S., {Malespin}, C.~A.,
  {Martin}, M., {Millan}, M., {Miller}, K., {Navarro-Gonzalez}, R., {Prats},
  B.~D., {Steele}, A., {Teinturier}, S., {Mahaffy}, P.~R., Mar. 2016. {First
  Detection of Non-Chlorinated Organic Molecules Indigenous to a Martian
  Sample}. In: Lunar and Planetary Science Conference. Vol.~47 of Lunar and
  Planetary Science Conference. p. 2568.

\bibitem[{{Freissinet} et~al.(2015){Freissinet}, {Glavin}, {Mahaffy}, {Miller},
  {Eigenbrode}, {Summons}, {Brunner}, {Buch}, {Szopa}, {Archer}, {Franz},
  {Atreya}, {Brinckerhoff}, {Cabane}, {Coll}, {Conrad}, {Des Marais},
  {Dworkin}, {Fair{\'e}n}, {Fran{\c c}ois}, {Grotzinger}, {Kashyap}, {ten
  Kate}, {Leshin}, {Malespin}, {Martin}, {Martin-Torres}, {McAdam}, {Ming},
  {Navarro-Gonz{\'a}lez}, {Pavlov}, {Prats}, {Squyres}, {Steele}, {Stern},
  {Sumner}, {Sutter}, {Zorzano}, and {MSL Science Team}}]{Freissinet2015}
{Freissinet}, C., {Glavin}, D.~P., {Mahaffy}, P.~R., {Miller}, K.~E.,
  {Eigenbrode}, J.~L., {Summons}, R.~E., {Brunner}, A.~E., {Buch}, A., {Szopa},
  C., {Archer}, Jr., P.~D., {Franz}, H.~B., {Atreya}, S.~K., {Brinckerhoff},
  W.~B., {Cabane}, M., {Coll}, P., {Conrad}, P.~G., {Des Marais}, D.~J.,
  {Dworkin}, J.~P., {Fair{\'e}n}, A.~G., {Fran{\c c}ois}, P., {Grotzinger},
  J.~P., {Kashyap}, S., {ten Kate}, I.~L., {Leshin}, L.~A., {Malespin}, C.~A.,
  {Martin}, M.~G., {Martin-Torres}, J.~F., {McAdam}, A.~C., {Ming}, D.~W.,
  {Navarro-Gonz{\'a}lez}, R., {Pavlov}, A.~A., {Prats}, B.~D., {Squyres},
  S.~W., {Steele}, A., {Stern}, J.~C., {Sumner}, D.~Y., {Sutter}, B.,
  {Zorzano}, M.-P., {MSL Science Team}, Mar. 2015. {Organic molecules in the
  Sheepbed Mudstone, Gale Crater, Mars}. Journal of Geophysical Research
  (Planets) 120, 495--514.

\bibitem[{{Glavin} et~al.(2013){Glavin}, {Freissinet}, {Miller}, {Eigenbrode},
  {Brunner}, {Buch}, {Sutter}, {Archer}, {Atreya}, {Brinckerhoff}, {Cabane},
  {Coll}, {Conrad}, {Coscia}, {Dworkin}, {Franz}, {Grotzinger}, {Leshin},
  {Martin}, {McKay}, {Ming}, {Navarro-Gonz{\'a}lez}, {Pavlov}, {Steele},
  {Summons}, {Szopa}, {Teinturier}, and {Mahaffy}}]{Glavin2013}
{Glavin}, D.~P., {Freissinet}, C., {Miller}, K.~E., {Eigenbrode}, J.~L.,
  {Brunner}, A.~E., {Buch}, A., {Sutter}, B., {Archer}, P.~D., {Atreya}, S.~K.,
  {Brinckerhoff}, W.~B., {Cabane}, M., {Coll}, P., {Conrad}, P.~G., {Coscia},
  D., {Dworkin}, J.~P., {Franz}, H.~B., {Grotzinger}, J.~P., {Leshin}, L.~A.,
  {Martin}, M.~G., {McKay}, C., {Ming}, D.~W., {Navarro-Gonz{\'a}lez}, R.,
  {Pavlov}, A., {Steele}, A., {Summons}, R.~E., {Szopa}, C., {Teinturier}, S.,
  {Mahaffy}, P.~R., Oct. 2013. {Evidence for perchlorates and the origin of
  chlorinated hydrocarbons detected by SAM at the Rocknest aeolian deposit in
  Gale Crater}. Journal of Geophysical Research (Planets) 118, 1955--1973.

\bibitem[{{Goesmann} et~al.(2015){Goesmann}, {Rosenbauer}, {Bredeh{\"o}ft},
  {Cabane}, {Ehrenfreund}, {Gautier}, {Giri}, {Kr{\"u}ger}, {Le Roy},
  {MacDermott}, {McKenna-Lawlor}, {Meierhenrich}, {Caro}, {Raulin}, {Roll},
  {Steele}, {Steininger}, {Sternberg}, {Szopa}, {Thiemann}, and
  {Ulamec}}]{Goesmann2015}
{Goesmann}, F., {Rosenbauer}, H., {Bredeh{\"o}ft}, J.~H., {Cabane}, M.,
  {Ehrenfreund}, P., {Gautier}, T., {Giri}, C., {Kr{\"u}ger}, H., {Le Roy}, L.,
  {MacDermott}, A.~J., {McKenna-Lawlor}, S., {Meierhenrich}, U.~J., {Caro},
  G.~M.~M., {Raulin}, F., {Roll}, R., {Steele}, A., {Steininger}, H.,
  {Sternberg}, R., {Szopa}, C., {Thiemann}, W., {Ulamec}, S., Jul. 2015.
  {Organic compounds on comet 67P/Churyumov-Gerasimenko revealed by COSAC mass
  spectrometry}. Science 349~(2).

\bibitem[{{Granvik} et~al.(2016){Granvik}, {Morbidelli}, {Jedicke}, {Bolin},
  {Bottke}, {Beshore}, {Vokrouhlick{\'y}}, {Delb{\`o}}, and
  {Michel}}]{Granvik2016}
{Granvik}, M., {Morbidelli}, A., {Jedicke}, R., {Bolin}, B., {Bottke}, W.~F.,
  {Beshore}, E., {Vokrouhlick{\'y}}, D., {Delb{\`o}}, M., {Michel}, P., Feb.
  2016. {Super-catastrophic disruption of asteroids at small perihelion
  distances}. \nat 530, 303--306.

\bibitem[{{Granvik} et~al.(2017){Granvik}, {Morbidelli}, {Vokrouhlick{\'y}},
  {Bottke}, {Nesvorn{\'y}}, and {Jedicke}}]{Granvik2017}
{Granvik}, M., {Morbidelli}, A., {Vokrouhlick{\'y}}, D., {Bottke}, W.~F.,
  {Nesvorn{\'y}}, D., {Jedicke}, R., Jan. 2017. {Escape of asteroids from the
  main belt}. \aap 598, A52.

\bibitem[{{Greenstreet} et~al.(2012){Greenstreet}, {Ngo}, and
  {Gladman}}]{Greenstreet2012}
{Greenstreet}, S., {Ngo}, H., {Gladman}, B., Jan. 2012. {The orbital
  distribution of Near-Earth Objects inside Earth's orbit}. \icarus 217,
  355--366.

\bibitem[{{Holsapple} and {Housen}(2007)}]{Holsapple2007}
{Holsapple}, K.~A., {Housen}, K.~R., Mar. 2007. {A crater and its ejecta: An
  interpretation of Deep Impact}. \icarus 187, 345--356.

\bibitem[{{Iglesias-Groth} et~al.(2011){Iglesias-Groth}, {Cataldo}, {Ursini},
  and {Manchado}}]{Iglesias-Groth2011}
{Iglesias-Groth}, S., {Cataldo}, F., {Ursini}, O., {Manchado}, A., Jan. 2011.
  {Amino acids in comets and meteorites: stability under gamma radiation and
  preservation of the enantiomeric excess}. \mnras 410, 1447--1453.

\bibitem[{{Jessberger} et~al.(1988){Jessberger}, {Christoforidis}, and
  {Kissel}}]{Jessberger1988}
{Jessberger}, E.~K., {Christoforidis}, A., {Kissel}, J., Apr. 1988. {Aspects of
  the major element composition of Halley's dust}. \nat 332, 691--695.

\bibitem[{{Kokubo} and {Ida}(1996)}]{Kokubo1996}
{Kokubo}, E., {Ida}, S., Sep. 1996. {On Runaway Growth of Planetesimals}.
  \icarus 123, 180--191.

\bibitem[{{Leshin} et~al.(2013){Leshin}, {Mahaffy}, {Webster}, {Cabane},
  {Coll}, {Conrad}, {Archer}, {Atreya}, {Brunner}, {Buch}, and
  et~al.}]{Leshin2013}
{Leshin}, L.~A., {Mahaffy}, P.~R., {Webster}, C.~R., {Cabane}, M., {Coll}, P.,
  {Conrad}, P.~G., {Archer}, P.~D., {Atreya}, S.~K., {Brunner}, A.~E., {Buch},
  A., et~al., Sep. 2013. {Volatile, Isotope, and Organic Analysis of Martian
  Fines with the Mars Curiosity Rover}. Science 341, 1238937.

\bibitem[{{Levison} and {Duncan}(1994)}]{Levison1994}
{Levison}, H.~F., {Duncan}, M.~J., Mar. 1994. {The long-term dynamical behavior
  of short-period comets}. \icarus 108, 18--36.

\bibitem[{{Levison} and {Duncan}(2013)}]{Levison2013}
{Levison}, H.~F., {Duncan}, M.~J., Mar. 2013. {SWIFT: A solar system
  integration software package}. Astrophysics Source Code Library.

\bibitem[{{Mahaffy} et~al.(2012){Mahaffy}, {Webster}, {Cabane}, {Conrad},
  {Coll}, {Atreya}, {Arvey}, {Barciniak}, {Benna}, {Bleacher}, {Brinckerhoff},
  {Eigenbrode}, {Carignan}, {Cascia}, {Chalmers}, {Dworkin}, {Errigo},
  {Everson}, {Franz}, {Farley}, {Feng}, {Frazier}, {Freissinet}, {Glavin},
  {Harpold}, {Hawk}, {Holmes}, {Johnson}, {Jones}, {Jordan}, {Kellogg},
  {Lewis}, {Lyness}, {Malespin}, {Martin}, {Maurer}, {McAdam}, {McLennan},
  {Nolan}, {Noriega}, {Pavlov}, {Prats}, {Raaen}, {Sheinman}, {Sheppard},
  {Smith}, {Stern}, {Tan}, {Trainer}, {Ming}, {Morris}, {Jones}, {Gundersen},
  {Steele}, {Wray}, {Botta}, {Leshin}, {Owen}, {Battel}, {Jakosky}, {Manning},
  {Squyres}, {Navarro-Gonz{\'a}lez}, {McKay}, {Raulin}, {Sternberg}, {Buch},
  {Sorensen}, {Kline-Schoder}, {Coscia}, {Szopa}, {Teinturier}, {Baffes},
  {Feldman}, {Flesch}, {Forouhar}, {Garcia}, {Keymeulen}, {Woodward}, {Block},
  {Arnett}, {Miller}, {Edmonson}, {Gorevan}, and {Mumm}}]{Mahaffy2012}
{Mahaffy}, P.~R., {Webster}, C.~R., {Cabane}, M., {Conrad}, P.~G., {Coll}, P.,
  {Atreya}, S.~K., {Arvey}, R., {Barciniak}, M., {Benna}, M., {Bleacher}, L.,
  {Brinckerhoff}, W.~B., {Eigenbrode}, J.~L., {Carignan}, D., {Cascia}, M.,
  {Chalmers}, R.~A., {Dworkin}, J.~P., {Errigo}, T., {Everson}, P., {Franz},
  H., {Farley}, R., {Feng}, S., {Frazier}, G., {Freissinet}, C., {Glavin},
  D.~P., {Harpold}, D.~N., {Hawk}, D., {Holmes}, V., {Johnson}, C.~S., {Jones},
  A., {Jordan}, P., {Kellogg}, J., {Lewis}, J., {Lyness}, E., {Malespin},
  C.~A., {Martin}, D.~K., {Maurer}, J., {McAdam}, A.~C., {McLennan}, D.,
  {Nolan}, T.~J., {Noriega}, M., {Pavlov}, A.~A., {Prats}, B., {Raaen}, E.,
  {Sheinman}, O., {Sheppard}, D., {Smith}, J., {Stern}, J.~C., {Tan}, F.,
  {Trainer}, M., {Ming}, D.~W., {Morris}, R.~V., {Jones}, J., {Gundersen}, C.,
  {Steele}, A., {Wray}, J., {Botta}, O., {Leshin}, L.~A., {Owen}, T., {Battel},
  S., {Jakosky}, B.~M., {Manning}, H., {Squyres}, S., {Navarro-Gonz{\'a}lez},
  R., {McKay}, C.~P., {Raulin}, F., {Sternberg}, R., {Buch}, A., {Sorensen},
  P., {Kline-Schoder}, R., {Coscia}, D., {Szopa}, C., {Teinturier}, S.,
  {Baffes}, C., {Feldman}, J., {Flesch}, G., {Forouhar}, S., {Garcia}, R.,
  {Keymeulen}, D., {Woodward}, S., {Block}, B.~P., {Arnett}, K., {Miller}, R.,
  {Edmonson}, C., {Gorevan}, S., {Mumm}, E., Sep. 2012. {The Sample Analysis at
  Mars Investigation and Instrument Suite}. Space Science Reviews 170,
  401--478.

\bibitem[{{Marchi} et~al.(2012){Marchi}, {Bottke}, {Kring}, and
  {Morbidelli}}]{Marchi2012}
{Marchi}, S., {Bottke}, W.~F., {Kring}, D.~A., {Morbidelli}, A., Apr. 2012.
  {The onset of the lunar cataclysm as recorded in its ancient crater
  populations}. Earth and Planetary Science Letters 325, 27--38.

\bibitem[{{Masiero} et~al.(2011){Masiero}, {Mainzer}, {Grav}, {Bauer}, {Cutri},
  {Dailey}, {Eisenhardt}, {McMillan}, {Spahr}, {Skrutskie}, {Tholen}, {Walker},
  {Wright}, {DeBaun}, {Elsbury}, {Gautier}, {Gomillion}, and
  {Wilkins}}]{Masiero2011}
{Masiero}, J.~R., {Mainzer}, A.~K., {Grav}, T., {Bauer}, J.~M., {Cutri}, R.~M.,
  {Dailey}, J., {Eisenhardt}, P.~R.~M., {McMillan}, R.~S., {Spahr}, T.~B.,
  {Skrutskie}, M.~F., {Tholen}, D., {Walker}, R.~G., {Wright}, E.~L., {DeBaun},
  E., {Elsbury}, D., {Gautier}, IV, T., {Gomillion}, S., {Wilkins}, A., Nov.
  2011. {Main Belt Asteroids with WISE/NEOWISE. I. Preliminary Albedos and
  Diameters}. \apj 741, 68.

\bibitem[{{Milani} and {Knezevic}(1990)}]{Milani1990}
{Milani}, A., {Knezevic}, Z., Dec. 1990. {Secular perturbation theory and
  computation of asteroid proper elements}. Celestial Mechanics and Dynamical
  Astronomy 49, 347--411.

\bibitem[{{Ming} et~al.(2014){Ming}, {Archer}, {Glavin}, {Eigenbrode}, {Franz},
  {Sutter}, {Brunner}, {Stern}, {Freissinet}, {McAdam}, and et~al.}]{Ming2014}
{Ming}, D.~W., {Archer}, P.~D., {Glavin}, D.~P., {Eigenbrode}, J.~L., {Franz},
  H.~B., {Sutter}, B., {Brunner}, A.~E., {Stern}, J.~C., {Freissinet}, C.,
  {McAdam}, A.~C., et~al., Jan. 2014. {Volatile and Organic Compositions of
  Sedimentary Rocks in Yellowknife Bay, Gale Crater, Mars}. Science 343,
  1245267.

\bibitem[{{Moore} and {Lewis}(1967)}]{Moore1967}
{Moore}, C.~B., {Lewis}, C.~F., Dec. 1967. {Total carbon content of ordinary
  chondrites}. Journal of Geophysical Research 72~(24).

\bibitem[{{Moores} and {Schuerger}(2012)}]{Moores2012}
{Moores}, J.~E., {Schuerger}, A.~C., Aug. 2012. {UV degradation of accreted
  organics on Mars: IDP longevity, surface reservoir of organics, and relevance
  to the detection of methane in the atmosphere}. Journal of Geophysical
  Research (Planets) 117, E08008.

\bibitem[{{Nesvorn{\'y}} et~al.(2011){Nesvorn{\'y}}, {Janches},
  {Vokrouhlick{\'y}}, {Pokorn{\'y}}, {Bottke}, and {Jenniskens}}]{Nesvorny2011}
{Nesvorn{\'y}}, D., {Janches}, D., {Vokrouhlick{\'y}}, D., {Pokorn{\'y}}, P.,
  {Bottke}, W.~F., {Jenniskens}, P., Dec. 2011. {Dynamical Model for the
  Zodiacal Cloud and Sporadic Meteors}. \apj 743, 129.

\bibitem[{{Pavlov} et~al.(2014){Pavlov}, {Pavlov}, {Ostryakov}, {Vasilyev},
  {Mahaffy}, and {Steele}}]{Pavlov2014}
{Pavlov}, A.~A., {Pavlov}, A.~K., {Ostryakov}, V.~M., {Vasilyev}, G.~I.,
  {Mahaffy}, P., {Steele}, A., Jun. 2014. {Alteration of the carbon and
  nitrogen isotopic composition in the Martian surface rocks due to cosmic ray
  exposure}. Journal of Geophysical Research (Planets) 119, 1390--1402.

\bibitem[{{Pavlov} et~al.(2012){Pavlov}, {Vasilyev}, {Ostryakov}, {Pavlov}, and
  {Mahaffy}}]{Pavlov2012}
{Pavlov}, A.~A., {Vasilyev}, G., {Ostryakov}, V.~M., {Pavlov}, A.~K.,
  {Mahaffy}, P., Jul. 2012. {Degradation of the organic molecules in the
  shallow subsurface of Mars due to irradiation by cosmic rays}. \grl 39,
  L13202.

\bibitem[{{Pravec} and {Harris}(2007)}]{Pravec2007}
{Pravec}, P., {Harris}, A.~W., Sep. 2007. {Binary asteroid population. 1.
  Angular momentum content}. \icarus 190, 250--259.

\bibitem[{{Sagan} and {Pollack}(1974)}]{Sagan1974}
{Sagan}, C., {Pollack}, J.~B., Apr. 1974. {Differential Transmission of
  Sunlight on Mars: Biological Implications}. \icarus 21, 490--495.

\bibitem[{{Sandford} et~al.(2006){Sandford}, {Al{\'e}on}, {Alexander}, {Araki},
  {Bajt}, {Baratta}, {Borg}, {Bradley}, {Brownlee}, {Brucato}, {Burchell},
  {Busemann}, {Butterworth}, {Clemett}, {Cody}, {Colangeli}, {Cooper},
  {D'Hendecourt}, {Djouadi}, {Dworkin}, {Ferrini}, {Fleckenstein}, {Flynn},
  {Franchi}, {Fries}, {Gilles}, {Glavin}, {Gounelle}, {Grossemy}, {Jacobsen},
  {Keller}, {Kilcoyne}, {Leitner}, {Matrajt}, {Meibom}, {Mennella},
  {Mostefaoui}, {Nittler}, {Palumbo}, {Papanastassiou}, {Robert}, {Rotundi},
  {Snead}, {Spencer}, {Stadermann}, {Steele}, {Stephan}, {Tsou}, {Tyliszczak},
  {Westphal}, {Wirick}, {Wopenka}, {Yabuta}, {Zare}, and
  {Zolensky}}]{Sandford2006}
{Sandford}, S.~A., {Al{\'e}on}, J., {Alexander}, C.~M.~O.~., {Araki}, T.,
  {Bajt}, S., {Baratta}, G.~A., {Borg}, J., {Bradley}, J.~P., {Brownlee},
  D.~E., {Brucato}, J.~R., {Burchell}, M.~J., {Busemann}, H., {Butterworth},
  A., {Clemett}, S.~J., {Cody}, G., {Colangeli}, L., {Cooper}, G.,
  {D'Hendecourt}, L., {Djouadi}, Z., {Dworkin}, J.~P., {Ferrini}, G.,
  {Fleckenstein}, H., {Flynn}, G.~J., {Franchi}, I.~A., {Fries}, M., {Gilles},
  M.~K., {Glavin}, D.~P., {Gounelle}, M., {Grossemy}, F., {Jacobsen}, C.,
  {Keller}, L.~P., {Kilcoyne}, A.~L.~D., {Leitner}, J., {Matrajt}, G.,
  {Meibom}, A., {Mennella}, V., {Mostefaoui}, S., {Nittler}, L.~R., {Palumbo},
  M.~E., {Papanastassiou}, D.~A., {Robert}, F., {Rotundi}, A., {Snead}, C.~J.,
  {Spencer}, M.~K., {Stadermann}, F.~J., {Steele}, A., {Stephan}, T., {Tsou},
  P., {Tyliszczak}, T., {Westphal}, A.~J., {Wirick}, S., {Wopenka}, B.,
  {Yabuta}, H., {Zare}, R.~N., {Zolensky}, M.~E., Dec. 2006. {Organics Captured
  from Comet 81P/Wild 2 by the Stardust Spacecraft}. Science 314, 1720.

\bibitem[{{Schuerger} et~al.(2003){Schuerger}, {Mancinelli}, {Kern},
  {Rothschild}, and {McKay}}]{Schuerger2003}
{Schuerger}, A.~C., {Mancinelli}, R.~L., {Kern}, R.~G., {Rothschild}, L.~J.,
  {McKay}, C.~P., Oct. 2003. {Survival of endospores of Bacillus subtilis on
  spacecraft surfaces under simulated martian environments: . implications for
  the forward contamination of Mars}. \icarus 165, 253--276.

\bibitem[{{Sephton}(2014)}]{Sephton2014}
{Sephton}, M.~A., Oct. 2014. {Organic Geochemistry of Meteorites}.

\bibitem[{{Sephton} et~al.(2002){Sephton}, {Wright}, {Gilmour}, {de Leeuw},
  {Grady}, and {Pillinger}}]{Sephton2002}
{Sephton}, M.~A., {Wright}, I.~P., {Gilmour}, I., {de Leeuw}, J.~W., {Grady},
  M.~M., {Pillinger}, C.~T., Jun. 2002. {High molecular weight organic matter
  in martian meteorites}. Planetary and Space Sciences Research Institute 50,
  711--716.

\bibitem[{{Speyerer} et~al.(2016){Speyerer}, {Povilaitis}, {Robinson},
  {Thomas}, and {Wagner}}]{Speyerer2016}
{Speyerer}, E.~J., {Povilaitis}, R.~Z., {Robinson}, M.~S., {Thomas}, P.~C.,
  {Wagner}, R.~V., Oct. 2016. {Quantifying crater production and regolith
  overturn on the Moon with temporal imaging}. \nat 538, 215--218.

\bibitem[{{Svetsov}(2011)}]{Svetsov2011}
{Svetsov}, V., Jul. 2011. {Cratering erosion of planetary embryos}. \icarus
  214, 316--326.

\bibitem[{{Svetsov} and {Shuvalov}(2015)}]{Svetsov2015}
{Svetsov}, V.~V., {Shuvalov}, V.~V., Nov. 2015. {Water delivery to the Moon by
  asteroidal and cometary impacts}. \planss 117, 444--452.

\bibitem[{{Swamy}(2010)}]{Swamy2010}
{Swamy}, K.~S.~K., 2010. {Physics of Comets (3rd Edition)}. World Scientific
  Publishing Co.

\bibitem[{{ten Kate}(2010)}]{tenKate2010}
{ten Kate}, I.~L., Aug. 2010. {Organics on Mars?} Astrobiology 10, 589--603.

\bibitem[{{ten Kate} et~al.(2005){ten Kate}, {Garry}, {Peeters}, {Quinn},
  {Foing}, and {Ehrenfreund}}]{tenKate2005}
{ten Kate}, I.~L., {Garry}, J.~R.~C., {Peeters}, Z., {Quinn}, R., {Foing}, B.,
  {Ehrenfreund}, P., Aug. 2005. {Amino acid photostability on the Martian
  surface}. Meteoritics and Planetary Science 40, 1185.

\bibitem[{{Turrini}(2014)}]{Turrini2014_model}
{Turrini}, D., Nov. 2014. {The primordial collisional history of Vesta: crater
  saturation, surface evolution and survival of the basaltic crust}. \planss
  103, 82--95.

\bibitem[{{Turrini} et~al.(2014){Turrini}, {Combe}, {McCord}, {Oklay},
  {Vincent}, {Prettyman}, {McSween}, {Consolmagno}, {De Sanctis}, {Le Corre},
  {Longobardo}, {Palomba}, and {Russell}}]{Turrini2014}
{Turrini}, D., {Combe}, J.-P., {McCord}, T.~B., {Oklay}, N., {Vincent}, J.-B.,
  {Prettyman}, T.~H., {McSween}, H.~Y., {Consolmagno}, G.~J., {De Sanctis},
  M.~C., {Le Corre}, L., {Longobardo}, A., {Palomba}, E., {Russell}, C.~T.,
  Sep. 2014. {The contamination of the surface of Vesta by impacts and the
  delivery of the dark material}. \icarus 240, 86--102.

\bibitem[{{Turrini} et~al.(2016){Turrini}, {Svetsov}, {Consolmagno}, {Sirono},
  and {Pirani}}]{Turrini2016}
{Turrini}, D., {Svetsov}, V., {Consolmagno}, G., {Sirono}, S., {Pirani}, S.,
  Dec. 2016. {Olivine on Vesta as exogenous contaminants brought by impacts:
  Constraints from modeling Vesta's collisional history and from impact
  simulations}. \icarus 280, 328--339.

\bibitem[{{Webster} et~al.(2015){Webster}, {Mahaffy}, {Atreya}, and
  {Flesch}}]{Webster2015}
{Webster}, C.~R., {Mahaffy}, P.~R., {Atreya}, S.~K., {Flesch}, G., Dec. 2015.
  {Mars Methane Detection and Variability at Gale Crater Measured by the TLS
  instrument in SAM on the Curiosity Rover}. AGU Fall Meeting Abstracts.

\bibitem[{{Wray}(2013)}]{Wray2013}
{Wray}, J.~J., Jan. 2013. {Gale crater: the Mars Science Laboratory/Curiosity
  Rover Landing Site}. International Journal of Astrobiology 12, 25--38.

\end{thebibliography}

\end{document}